\documentclass[%
reprint,
superscriptaddress,
amsmath,amssymb,
aps,
prl,
floatfix
]{revtex4-2}

\usepackage{graphicx}% Include figure files
\usepackage{dcolumn}% Align table columns on decimal point
\usepackage{bm}% bold math
\usepackage{comment}% 
\usepackage{amsmath}% 
\usepackage{url}% 
\usepackage[mathlines,pagewise]{lineno}% Enable numbering of text and display math
\begin{document}
	
	\preprint{APS/123-QED}
	
	\title{Transient Rheology of Immersed Granular Materials}% Force line breaks with \\
	\author{Zhuan Ge}
    \affiliation{Zhejiang University, 866 Yuhangtang Road, Hangzhou 310058, Zhejiang, China\\}
\affiliation{ Key laboratory of Coastal Environment and Resources of Zhejiang Province (KLaCER), School of Engineering, Westlake University, 18 Shilongshan Street, Hangzhou, Zhejiang 310024, China.\\}

	\author{Teng Man}%
	\email{manteng@westlake.edu.cn}
\affiliation{ Key laboratory of Coastal Environment and Resources of Zhejiang Province (KLaCER), School of Engineering, Westlake University, 18 Shilongshan Street, Hangzhou, Zhejiang 310024, China.\\}
\author{Herbert E. Huppert}%
\affiliation{
	Institute of Theoretical Geophysics, King's College, University of Cambridge, King's Parade, Cambridge CB2 1ST, United Kingdom\\
}%
\author{Sergio Andres Galindo-Torres}
\email{s.torres@westlake.edu.cn}
\affiliation{ Key laboratory of Coastal Environment and Resources of Zhejiang Province (KLaCER), School of Engineering, Westlake University, 18 Shilongshan Street, Hangzhou, Zhejiang 310024, China.\\}
\date{\today}% It is always \today, today,
\begin{abstract}
	In this letter, we investigate the transient rheological behavior of immersed granular flows using both experiments of submerged granular column collapses and corresponding numerical simulations. The simulations are performed with the lattice-Boltzmann method (LBM) coupled with the discrete element method (DEM) and provide a significant amount of data of the stress and deformation conditions at different positions and times during the granular collapse. We derive a new dimensionless number $\mathcal{G}$ that can unify the rheology of transient granular flows in different regimes for all the simulation data points. $\mathcal{G}$ smoothly transforms from an inertial number into a viscous number, unifying both extremes of the rheology law. We also show the need to introduce the kinetic stresses to achieve a universal relation. The findings establish a transient constitutive framework for visco-inertial granular flows, and are important for a better understanding of granular-fluid mixtures in both natural and engineering situations. 
	
\end{abstract}

\maketitle

%\tableofcontents

$Introduction-$Granular flows are ubiquitous in natural phenomena, such as landslides, debris flows, and rock falls \cite{1995The,2002Avalanche,2020Pore}, and they can exhibit different flow behaviors akin to solids, fluids, or gases \cite{jaeger1996granular,roux2002quasistatic,goldhirsch2003rapid,GDR2004On}. Complex environmental conditions, transient fluctuations, and highly dissipative interactions make it difficult to obtain a unified constitutive law for their flow characteristics. 
Following pioneering works \cite{iordanoff2004granular,da2005rheophysics,jop2006constitutive} on dry granular flows in steady state conditions, it has been determined that the apparent frictional coefficient $\mu$ can be considered a sole function of the inertial number $I$. This inertial number $I=\dot{\gamma}d/\sqrt{P/\rho_{s}}$ is defined as the ratio of a microscopic time scale ($\sqrt{d^{2}\rho_{s}/P}$) to a macroscopic deformation time scale ($1/\dot{\gamma}$) \cite{GDR2004On,da2005rheophysics}, where $\dot{\gamma}$ is the shear strain rate, $d$ is the averaged particle diameter, $P$ is the pressure applied to the granular sample, and $\rho_{s}$ is the particle density. Lacaze et al.\cite{lacaze2009axisymmetric} verified this theory through transient granular column collapse experiments, showing a successful application of the $\mu(I)$ theory to granular flows.

At the other end of the spectrum, in the case of fully submerged granular flows, Boyer et al.\cite{boyer2011unifying} showed the apparent friction $\mu(I_{v})$ of dense suspensions is a function of the viscous number $I_{v}=\eta_{f}\dot{\gamma}/P$, where $\eta_{f}$ is the fluid dynamic viscosity. Trulsson et al.\cite{trulsson2012transition} further investigated the rheology of submerged granular flow in the visco-inertial regime, and proposed a combined dimensionless number $K=\lambda I^{2}+I_{v}$ for successfully describing the submerged granular flows in different flow regimes (by varying the viscosity of interstitial fluid). However, $\lambda$ is obtained by fitting and has no clear physical definition. Furthermore, the work of Lacaze et al \cite{lacaze2021immersed} showed that $K$ cannot describe the transient rheology of immersed granular flows \cite{du2003granular} with enough accuracy. Although previous works improve the understanding of granular flows in diverse conditions, there is still work to be done to translate these models into predictive tools for natural hazards \cite{forterre2008flows}. More effort is needed to establish a universal constitutive law suitable for the complex granular flow where both particle interactions and hydrodynamic forces are non-negligible. In this case, where the granular assembly goes from a granular skeleton (where friction rules) to a dense suspension (where the viscosity of the fluid offers the greatest shear resistance), a proper rheology law is still lacking. 

Inspired by the complex dynamics presented in granular column collapses reported in Refs.\cite{thompson2007granular,bougouin2018granular,2020Pore}, we establish a numerical model using the Lattice-Boltzmann method (LBM) coupled with the discrete element method (DEM) to describe the immersed granular system and study the transient granular rheology based on granular column collapses. The numerical model is validated by immersed granular column collapse experiments. The one-to-one comparison between experiments and numerical simulations 
gives significant data for up-scaling the microscopic mechanism into macroscopic constitutive behaviors. Then, the rheology of immersed granular flow is investigated for systems in viscous, inertial, and free-fall regimes. The three flow regimes \cite{du2003granular}, which depend on the square root of the grain/fluid density ratio $r=\sqrt{\rho_{s}/\rho_{l}}$ and the Stokes number $St=[\rho_{s}(\rho_{s}-\rho_{l})gd^{3}]^{1/2}/(18\sqrt{2}\eta_{f})$, are classified to describe the effect of the fluid on grains in submerged granular flows.

\begin{figure}
	\centering
	\includegraphics[scale=0.23]{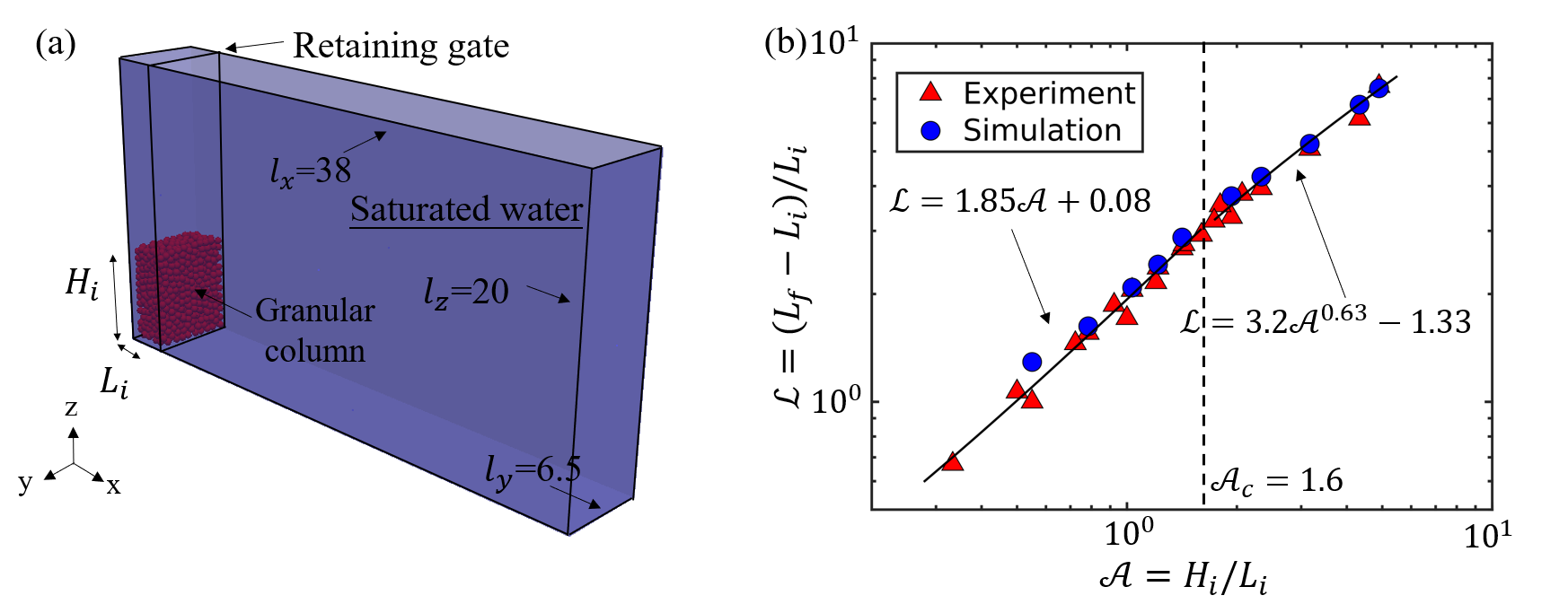}
	\caption{(a) Experimental setup of the submerged granular column collapse. (b) Dimensionless runout length ($\mathcal{L}=(L_{f}-L_{i})/L_{i}$) as a function of the aspect ratio $\mathcal{A}=H_{i}/L_{i}$ of granular column collapse in fluid for experiments and simulation results, $\mathcal{A}_{c}=1.6$ is the transition point.}
	\label{SGCsetup}
\end{figure}

$Experimental\,setup-$As presented in Fig. \ref{SGCsetup}, the dimension of the transparent plastic tank is 38 cm$\times$6.5 cm$\times$20 cm. Three positions are considered for the vertical retaining gate, corresponding to three different initial column lengths $L_{i}=$ 3, 6, and 9 cm to generate different sizes of initial granular columns. Plastic beads are used in this study. Their density is 1.18 g/cm$^3$, frictional coefficient is 0.34$\pm$0.01, and radius is 0.245$\pm$0.004 cm. Each test is recorded by a high-resolution camera with a frame rate of 100 fps. The particles are immersed in water for which the dynamic viscosity is 0.001 Pa${\cdot}$s and the density is 1 g/cm$^3$.
First, the retaining wall is placed at the desired position. Plastic particles are then gently poured into the reservoir delimited by the wall to generate the initial granular column, after which we pour the liquid into the tank until it reaches the desired level. Once the fluid surface and particles are static, we measure the initial length $L_{i}$ and initial height $H_{i}$ of the granular column. Then, the retaining wall is removed suddenly, and the column collapses and propagates into the tank. When particles stop propagating, we measure the deposit length, $L_{f}$, which is the final front position, and the final peak height, $H_{f}$. In this work, the initial aspect ratio, $\mathcal{A}=H_{i}/L_{i}$, of the granular column is varied within the range of 0.3-5. More details about the experiments are presented in the Supplementary Material.

\begin{figure}
	\centering
	\includegraphics[scale=0.4]{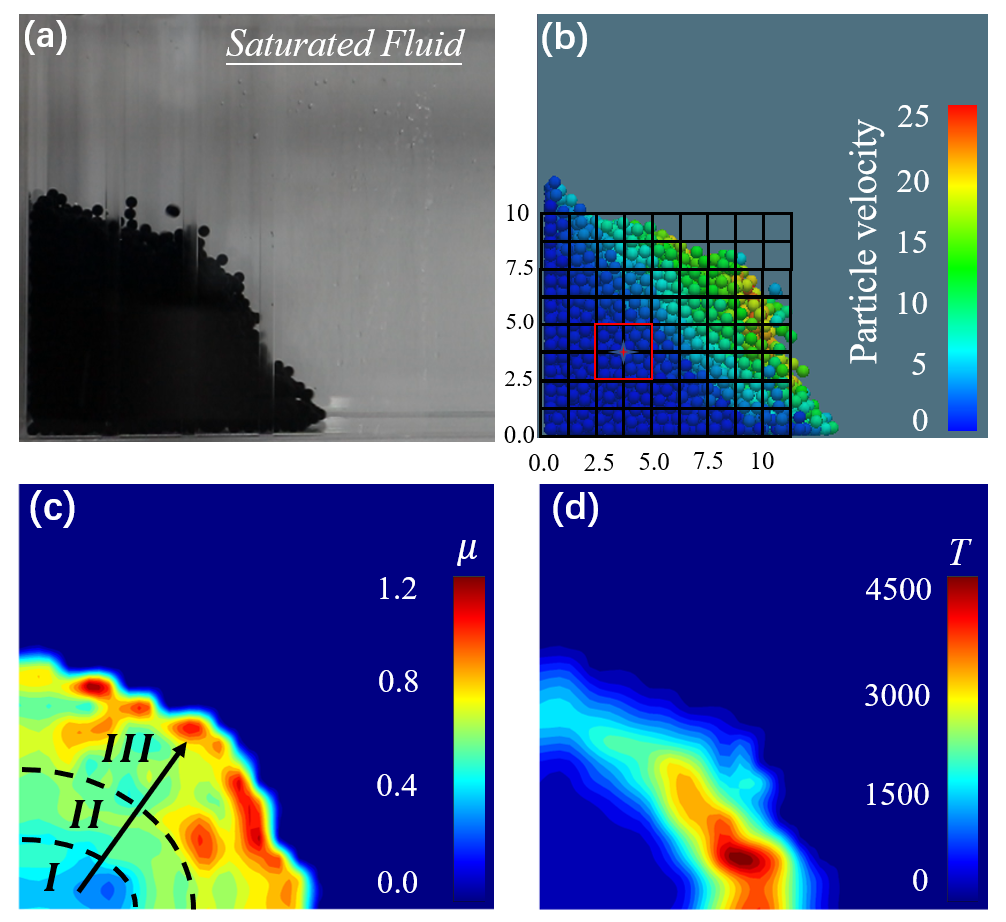}
	\caption{Submerged granular column collapse in water with a aspect ratio of 1.73 at 0.3s: (a) Experiment. (b) Numerical simulation. (c) Apparent frictional coefficient ($\mu$) distribution at y = 3 cm, area $I$, $\mu \le 0.4$, area $II$; $0.4< \mu \le 0.6$, area $III$; $\mu \ge 0.6$. (d) Granular temperature ($T$) distribution.}
	\label{shot}
\end{figure}

$Simulation\,setup-$We use DEM with frictional contact interactions modeled by a Hookean contact law with energy dissipation \cite{cundall1979discrete}. LBM is used to simulate the fluid flow in the pore space and to calculate the momentum exchange between the fluid and the particles \cite{galindo2013coupled}.
As shown in Fig. \ref{SGCsetup}(b), the normalized run-out distance $\mathcal{L}=(L_{f}-L_{i})/L_{i}$ shows good agreement between experiments and numerical simulations. The transition point appears when the aspect ratio is at $\mathcal{A}_{c}=$1.6, which is similar to Ref.\cite{bougouin2018granular}, and smaller than the granular collapse in dry conditions \cite{lube2005collapses}.  
During the granular collapse process, shown in Fig. \ref{shot}, the profile of the granular assembly also shows good agreement. 

\begin{figure}
	\centering
	\includegraphics[scale=0.3]{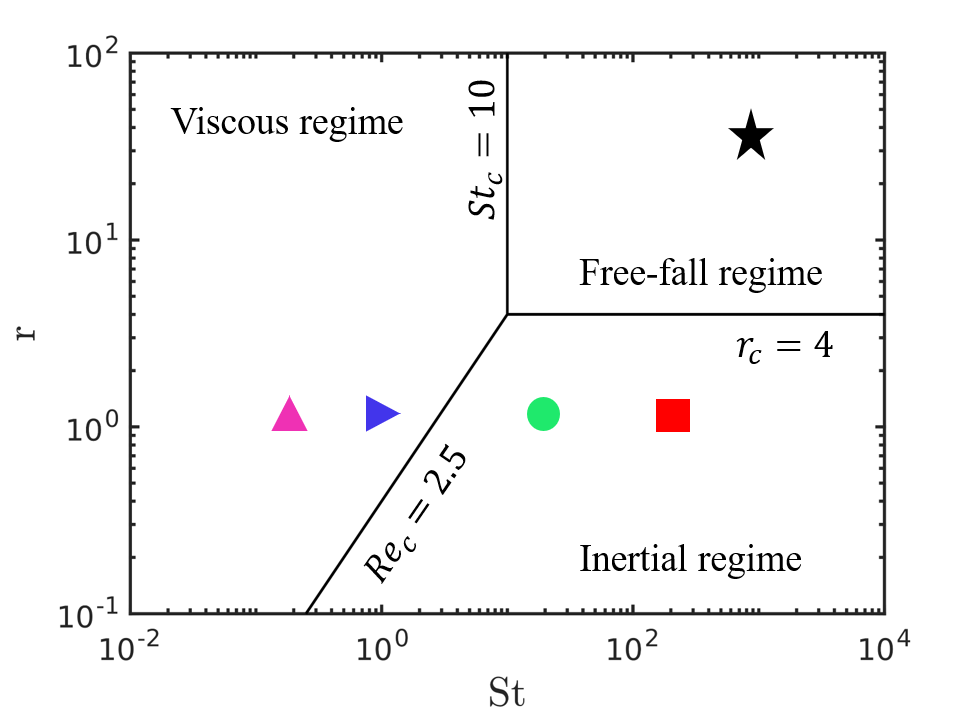}
	\caption{Granular-fluid flow regimes in the (St,r) plane according to Ref.\cite{du2003granular} implemented in the simulation: dry condition (black $\bigstar$ in the free-fall regime), immersed condition with different fluid viscosity (pink $\bigtriangleup$, $\eta_{f}=$1 g/(cm$\cdot$s) and dark blue $\rhd$, $\eta_{f}=$0.2 g/(cm$\cdot$s) in the viscous regime; the green $\bigcirc$, $\eta_{f}=$0.01 g/(cm$\cdot$s) and red $\square$, $\eta_{f}=$0.001 g/(cm$\cdot$s) in inertial regime.)}
	\label{STR}
\end{figure}

We implement immersed granular column collapse simulations at the initial aspect ratio $\mathcal{A}$=1.73 (the initial height is 10.3 cm and the initial length is 6 cm) with different viscosities as shown in Fig. \ref{STR} to investigate their rheological behaviors in different flow regimes. 
The granular system is then discretized into several representative volume elements (RVE) as shown in Fig. \ref{shot}(b) as a black grid, with side 1 cm, the particles in the surrounding four cubics of each cell are used to obtain the macroscopic information such as the averaged stress $\sigma$, strain rate $\dot{\gamma}$, solid fraction $\phi$, and granular temperature $T=\delta v^2/D$, where $\delta v$ is the velocity fluctuation, and $D$ is the space dimension. The averaged stress is calculated by the contact term $\sigma_{c}=\frac{1}{V}\sum_{p\in V}f_{i}l_{j}$, where $f_{i}$ is the $i$ component of the contact force between colliding DEM particles, and $l_{j}$ is the $j$ component of the branch vector, and $i,j$ represents the $x, y, z$ direction. This contact stress tensor needs to be corrected to account for small REVs as discussed in Ref.\cite{yan2019definition}. However, although this tensor is widely used for DEM studies, it will be shown later how it must be complemented by the kinetic stress tensor to achieve a universal rheology law.  The pressure $P$ and the shear stress $\tau$ are given by $P=-(\sigma_{11}+\sigma_{22}+\sigma_{33})/3$ and $\tau=\sqrt{1/2\tau_{ij}\tau_{ij}}$, respectively, where $\tau_{ij}=\sigma_{ij}+P \delta_{ij}$ is the deviatoric stress tensor. The equivalent strain rate tensor is calculated through the coarse-graining approach as described in Ref.\cite{goldhirsch2002microscopic}. 

\begin{figure}
	\centering
	\includegraphics[scale=0.21]{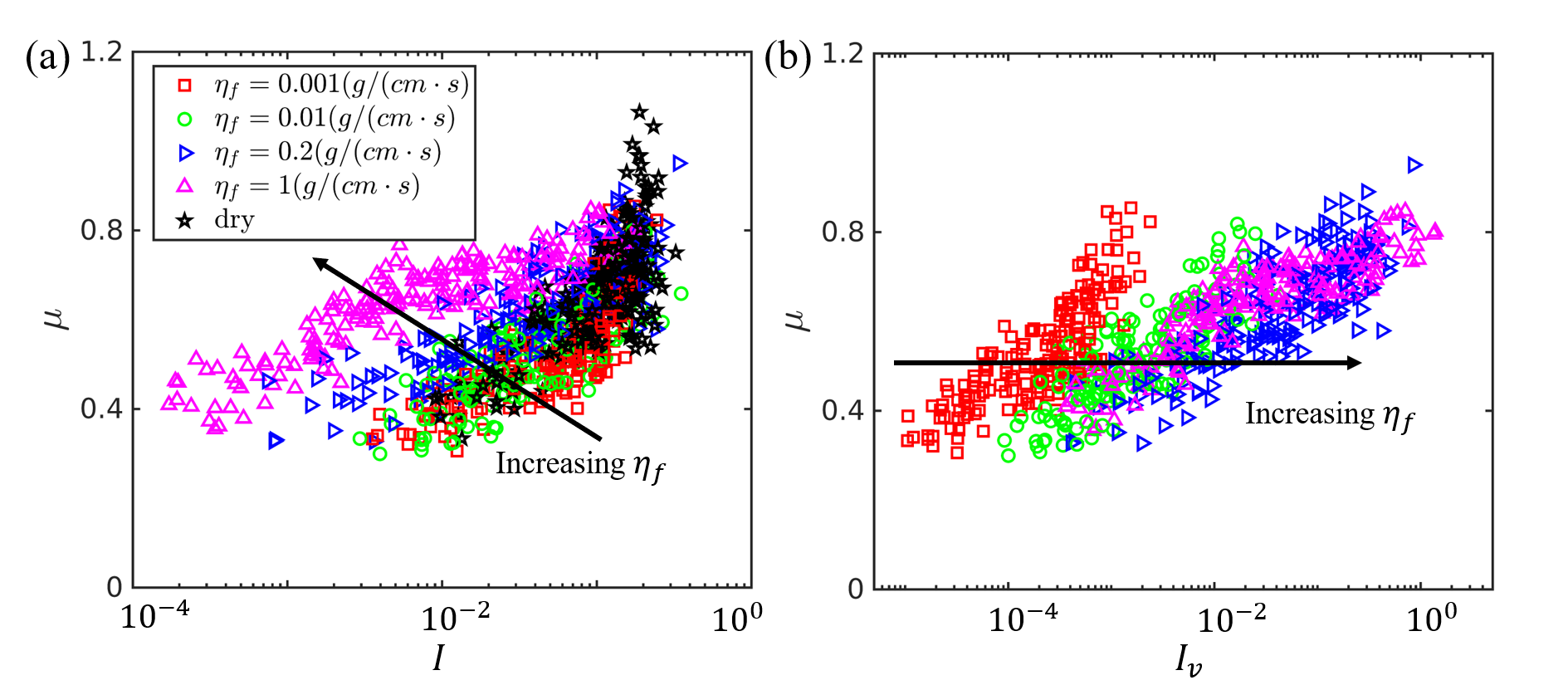}
	\caption{(a) Relationship between $\mu$ and $I$. (b) Relationship between $\mu$ and $I_{v}$. Data are obtained from the beginning to end of granular collapse in fluid with different viscosities.}
	\label{MuT}
\end{figure}

As shown in Fig. \ref{shot}(c), the apparent frictional coefficient increases spatially along the arrow direction. When the granular collapse is in the dense quasi-static regime (area $I$), the apparent frictional coefficient is close to the microscopic frictional coefficient, where $\mu \le 0.4$. In the area $II$, with increasing of granular velocity, the apparent frictional coefficient increases, where $0.4 < \mu \le 0.6$. Larger frictional coefficient appear associated with large granular temperatures near the interface between the granular material and fluid (area $III$), where $\mu > 0.6$. This is clear by comparing area $III$ with the region of high temperature shown in Fig. \ref{shot}(d). We consider this as a hint to introduce the kinetic stress tensor in our analysis as will be shown later. In Fig. \ref{MuT}(a), we plot the relationship between $\mu$ and $I$, where we find that the $\mu-I$ rheology is sufficient to describe the constitutive relationship of systems in inertial regimes [dry, $\eta_{f}=0.001$ g/(cm$\cdot$s), $\eta_{f}=0.01$ g/(cm$\cdot$s)]. However, as we increase the fluid viscosity, to reach the viscous regime, the $\mu-I$ relationship of systems with $\eta_{f}=0.2$ g/(cm$\cdot$s) and $\eta_{f}=1$ g/(cm$\cdot$s) deviates from the others. Plotting the relationship between $\mu$ and $I_{v}$ in Fig. \ref{MuT}(b) shows that the data have a better collapse for systems with $\eta_{f}=0.2$ g/(cm$\cdot$s) and $\eta_{f}=1$ g/(cm$\cdot$s), but cannot capture the behavior of systems with $\eta_{f}=0.01$ and $\eta_{f}=0.001$ and the dry sample. Systems in different regimes result in distinct rheological behaviors, which further indicates that a universal rheology is needed to describe the transient granular flow among free-fall, inertial, and viscous regimes.

\begin{figure}
	\centering
	\includegraphics[scale=0.22]{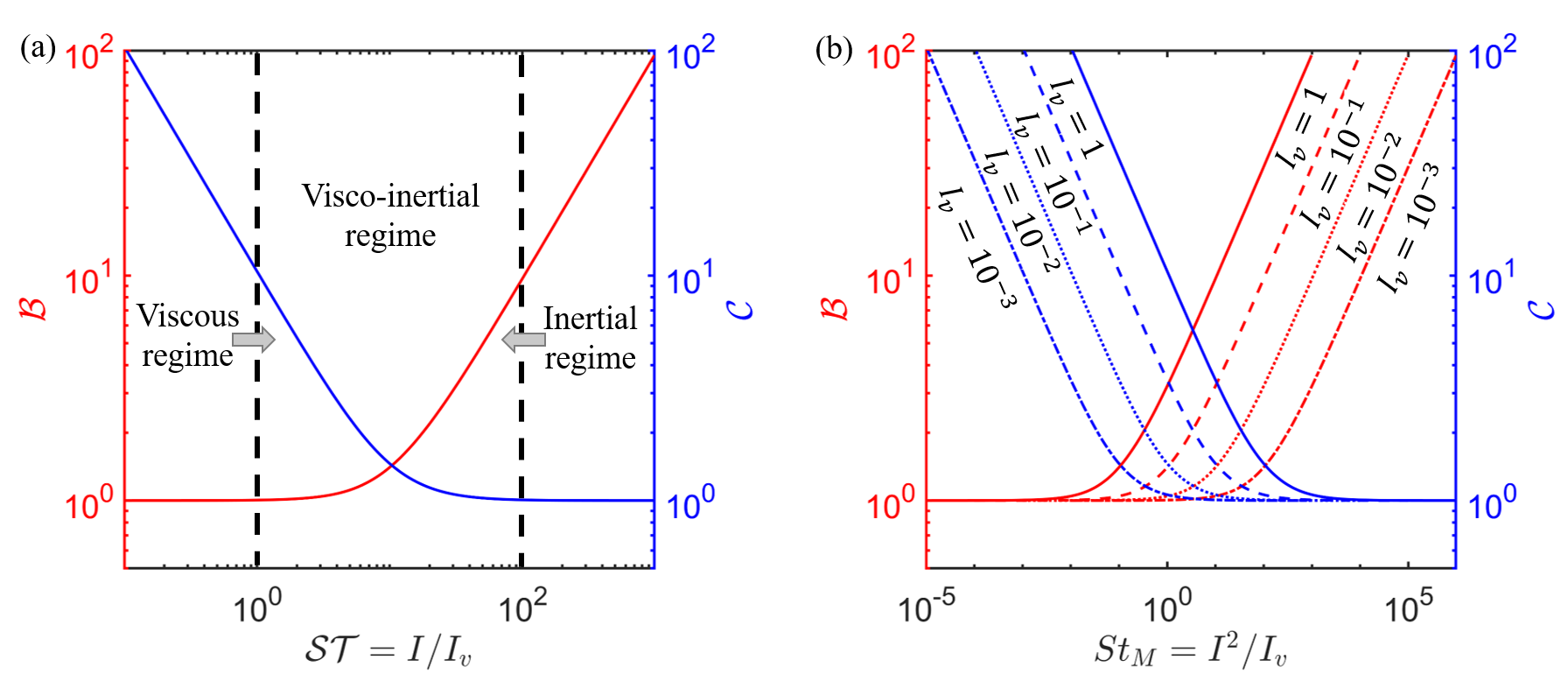}
	\caption{Evolution of $\mathcal{B}=\frac{\mathcal{G}}{12I_{v}}$ and $\mathcal{C}=\frac{\mathcal{G}}{2I/\sqrt{3}}$ in terms of (a)  $\mathcal{ST}=I/I_{v}$, (b) $St_{M}=I^2/I_{v}$.}
	\label{bta}
\end{figure}

The rheology of the granular flow is usually represented by the microscopic particle movement time scale $t_{f}$ divided by the macroscopic rearrangement time scale $\mathcal{T}=1/\dot{\gamma}$, where $t_{f}$ can be seen as the time for a particle to travel over a characteristic length, e.g. the particle diameter, $d$.
\begin{figure*}
	\centering
	\includegraphics[scale=0.30]{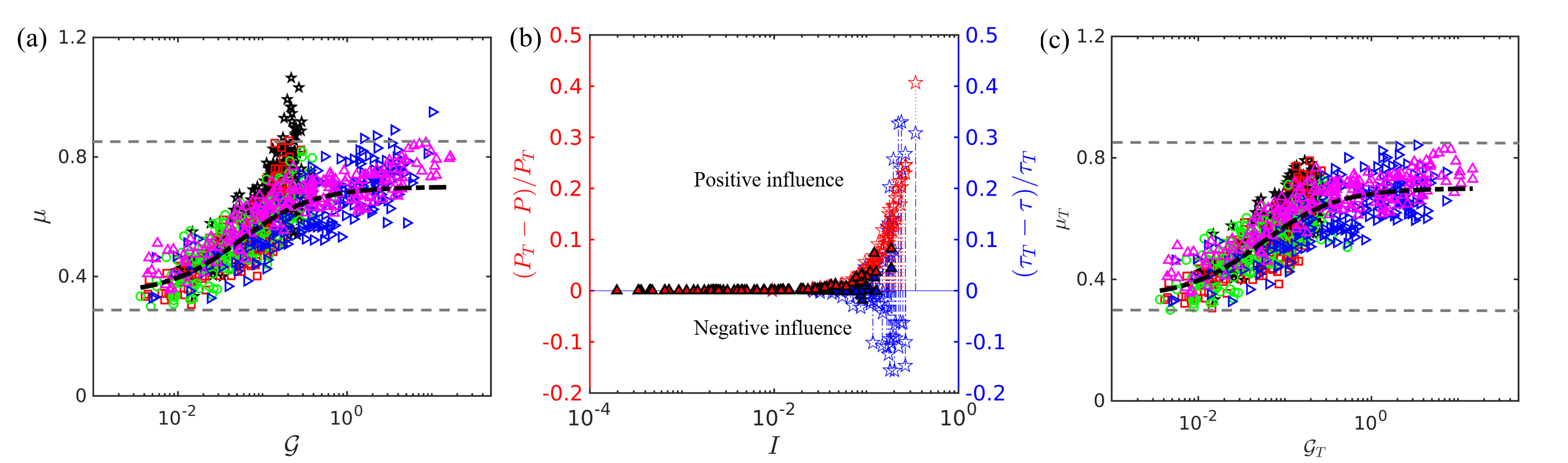}
	\caption{(a) $\mu$ as a function of $\mathcal{G}$ for different times in fluids of different viscosities. The symbols are for the same simulation cases with the same symbols coding in Figure \ref{MuT}. (b) Plots of the relative kinetic effect, ($(P_{T}-P)/P_{T}$ in the normal direction, color of blue, and $(\tau_{T}-\tau)/\tau_{T}$ in the tangential direction, color of red), with respect to the inertial number, $I$, for both dry (symbol $\bigstar$) and submerged (fluid viscosity is 1 g/(cm$\cdot$s), symbol $\bigtriangleup$) cases. (c) Apparent frictional coefficient that includes the kinetic effect $\mu_{T}$ as a function of $\mathcal{G}_{T}$ (incorporating kinetic effects).}
	\label{MuN}
\end{figure*}
In an ideal condition, the equilibrium of a single particle settling in the fluid is given by $(\pi/6)\rho_{s}d^3\frac{d u_{p}}{d t}=(\pi/4) P d^2-F_{d}$, where $F_{d}$ is the hydrodynamic force in submerged condition \cite{cassar2005submarine} and $u_{p}$ is the particle velocity. In the granular flow, the Reynolds number is usually very low, hence the hydrodynamic force can be assumed as the Stokes force $F_{d}=3\pi\eta_{f}d u_{p}$. During the settling process of a single particle, the particle velocity increases until the hydrodynamic force is equivalent to the inertial force, and the particle reaches the maximal final velocity $u_{f}=Pd/(12\eta_{f})$. In previous works \cite{jop2006constitutive,boyer2011unifying}, at the inertial regime (or dry granular flow), the drag force is neglected, which assumes that particles travel with a constant acceleration $a_{c}=3P/(2\rho_{s} d)$, with a deduced settling time $t_{f}=t_{dry}=\sqrt{4\rho_{s}d^2/(3P)}$ and a time scale ratio $t_{f}/\mathcal{T}=(2/\sqrt{3})I$. The constant factor $2/\sqrt{3}$ is usually ignored. In the viscous regime, it is assumed that the particle travels with the maximum velocity $u_{f}=Pd/(12\eta_{f})$ for the characteristic length $d$, the settling time is $t_{f}=t_{sub}=12\eta_{f}/P$, and the time scale ratio $t_{f}/\mathcal{T}=12I_{v}$. However, when the particle flows in the fluid where the inertial force is comparable to the hydrodynamic force, also known as the viscous-inertial regime, the rheology can be described by neither the inertial number $I$ nor viscous number $I_{v}$ individually. Hence, we derive the travel time in a transient condition as shown in the Appendix, where $t_{f}=\frac{12\eta_{f}}{P}+\frac{\rho_{s}d^2}{18\eta_{f}}[1-e^{-\frac{36\eta_{f}}{\sqrt{3P\rho_{s}}d}}]$, and obtain the ratio between microscopic time scale and macroscopic rearrangement time scale as
\begin{equation}
    \begin{split}
    \mathcal{G}=\frac{t_{f}}{\mathcal{T}}&=12I_{v}+\frac{I^2}{18 I_{v}}\left [1-e^{-\frac{36}{\sqrt{3}}\frac{I_{v}}{I}}\right ]\\
    &=12I_{v}\left [1+\frac{\mathcal{ST}^2}{216}-\frac{\mathcal{ST}^2}{216}e^{-\frac{36}{\sqrt{3}\mathcal{ST}}}\right ]\\
    &=\frac{2I}{\sqrt{3}}\left [\frac{6\sqrt{3}}{\mathcal{ST}}+\frac{\sqrt{3}\mathcal{ST}}{36}-\frac{\sqrt{3}\mathcal{ST}}{36}e^{-\frac{36}{\sqrt{3}\mathcal{ST}}}\right ],
    \end{split}
\end{equation}

where $\mathcal{ST}=I/I_{v}$. We define $\mathcal{B}= 1+\mathcal{ST}^2/216[1-e^{-36/(\sqrt{3}\mathcal{ST})}]$ and $\mathcal{C}= 6\sqrt{3}/\mathcal{ST}+\sqrt{3}\mathcal{ST}/36[1-e^{-36/(\sqrt{3}\mathcal{ST})}]$, so that $\mathcal{G}=\mathcal{B}(12I_{v})$ or $\mathcal{G}=\mathcal{C}(2I/\sqrt{3})$. Increasing the Stokes number leads to $\lim_{\mathcal{ST}\to \infty}\mathcal{G}=2I/\sqrt{3}=t_{dry}/\mathcal{T}$, while decreasing $\mathcal{ST}$ leads to $\lim_{\mathcal{ST}\to 0}\mathcal{G}=12I_{v}=t_{sub}/\mathcal{T}$. Hence, as granular materials flow from a free fall regime to a viscous regime, the dimensionless number $\mathcal{G}$ naturally transforms from an inertial number to a viscous number. Furthermore, the transition of flow regime of granular assemblies is redefined as shown in Fig.\ref{bta}(a). For $\mathcal{ST}\le1$, the granular flow is characterized by viscous number individually: this is the viscous regime. For $\mathcal{ST}>100$, the inertial number should be used instead: this is the inertial regime. For $1<\mathcal{ST}\le100$, the granular flow in the fluid is characterized by both viscous number and inertial number: this is the visco-inertial regime. However, the modified Stokes number $St_{M}=I^{2}/I_{v}$ cannot obtain a universal flow regime transition under different conditions,  as shown in Fig.\ref{bta}(b). These analytical results are consistent with the work of Trulsson et al.\cite{trulsson2012transition}, where they present the fraction of the power dissipated by each force field such as the inertial forces and hydrodynamic forces with different $\mathcal{ST}$. The hydrodynamic force is dominant when $\mathcal{ST}\le1$, while the contact force (inertial force) is dominant when $\mathcal{ST}>100$. Owing to $\mathcal{G}=12[I_{v}+I^{2}(1-e^{-36I_{v}/(\sqrt{3}I)})/(216I_{v})]$, the factor in dimensionless number $K=I _{v}+\lambda I^{2}$ is $\lambda=(1-e^{\frac{-36I_{v}}{\sqrt{3}I}})/(216I_{v})$, which depends on $I$ and $I_{v}$. This accounts for the reason why $\lambda$ varies in different cases\cite{trulsson2012transition,lacaze2021immersed,tapia2022viscous}.

As shown in Fig. \ref{MuN}(a), the rheology of immersed granular column collapses in different flow regimes can be partly described by the dimensionless number $\mathcal{G}$. With an increasing inertial number, the apparent frictional coefficient increases dramatically, as shown in Fig. \ref{MuT}(a), which results in the divergence of the relation between $\mathcal{G}$ and $\mu$ in Fig. \ref{MuN}(a), which is significant for the case of dry grains. We include the dynamic effect, to account for this inertia, by introducing the kinetic stress tensor $\sigma_{k}$. The total stress $\sigma_{T}=\sigma_{k}+\sigma$
is the combination of the kinetic part $\sigma_{k}=\frac{1}{V}\sum_{p\in V}m^{p}\delta v_{i}^{p}\delta v_{j}^{p}$ \cite{todd1995pressure,subramaniyan2008continuum} and the potential part $\sigma$. To explain why introducing the total stress tensor works, we introduce the following quantities: $\mathcal{G}_{T}$, and $\mu_{T}=\tau_{T}/P_{T}$ are calculated from the total stress $\sigma_{T}$, $\tau_{T}$ and $P_{T}$ are the shear stress and pressure derived from the total stress. When the inertial number is large, the kinetic stress increases the pressure (resulting in a positive $P_T-P$ difference), while reducing the shear stress (negative $\tau_T-\tau$), as shown in Fig. \ref{MuN}(b). The decrease in pressure $P$ and increase in shear stress $\tau$ results in an ever increasing value for $\mu$. Once the total stress tensor is considered, such discrepancy disappears for the total friction coefficient $\mu_T$. Furthermore, as shown in Fig. \ref{MuN}(c), $\mu_T$ for both dry and submerged conditions with a broad range of different viscosities could be generalized as a function of $\mathcal{G}_{T}$, with the following functional form
\begin{equation}
\mu_{T}(\mathcal{G}_{T})=\mu_{c}+\frac{\mu_{F}-\mu_{c}}{1+\mathcal{G}_{0}/\mathcal{G}_{T}}\label{Gn},
\end{equation}
where $\mu_{c}=0.34$ is the minimum apparent frictional coefficient, $\mu_{F}=0.7$ is the maximal macroscopic frictional coefficient, and $\mathcal{G}_{0}=0.054$ is a fitting factor.
In addition, Eq.\ref{Gn} can be used to quantify transient granular flows and predict their natural phenomena for the complex systems often observed in debris flows and submarine landslides.

$Conclusion-$Using DEM-LBM simulations, we have proposed a general constitutive relationship that is suitable for granular flow in different conditions. The new dimensionless number $\mathcal{G}$ can accurately describe the granular material flow in fluid, where the effect of fluid and grain force changes under different confining pressures, fluid viscosities, and macroscopic deformations. $\mathcal{G}$ naturally transforms into $I$, when hydrodynamic effects are negligable, and it converges to $I_{v}$ when those effects are significant.  It is shown how the kinetic stress, which is essential for transient granular systems, must be introduced into the rheological relationship in order to achieve universality. The proposed rheology law could be used to formulate constitutive models for large scale prediction at larger scales than the ones explored in this study.

This work is supported by the National Natural Science Foundation of China (NSFC major project grant NO. 12172305). We thank Westlake University Supercomputer Center for computational resources and related assistance. The simulations were based on the MECHSYS open source library (\url{http://mechsys.nongnu.org}).

\nocite{*}

\bibliography{apssamp}% Produces the bibliography via BibTeX.

\providecommand{\noopsort}[1]{}\providecommand{\singleletter}[1]{#1}%
\begin{thebibliography}{25}%
\makeatletter
\providecommand \@ifxundefined [1]{%
 \@ifx{#1\undefined}
}%
\providecommand \@ifnum [1]{%
 \ifnum #1\expandafter \@firstoftwo
 \else \expandafter \@secondoftwo
 \fi
}%
\providecommand \@ifx [1]{%
 \ifx #1\expandafter \@firstoftwo
 \else \expandafter \@secondoftwo
 \fi
}%
\providecommand \natexlab [1]{#1}%
\providecommand \enquote  [1]{``#1''}%
\providecommand \bibnamefont  [1]{#1}%
\providecommand \bibfnamefont [1]{#1}%
\providecommand \citenamefont [1]{#1}%
\providecommand \href@noop [0]{\@secondoftwo}%
\providecommand \href [0]{\begingroup \@sanitize@url \@href}%
\providecommand \@href[1]{\@@startlink{#1}\@@href}%
\providecommand \@@href[1]{\endgroup#1\@@endlink}%
\providecommand \@sanitize@url [0]{\catcode `\\12\catcode `\$12\catcode
  `\&12\catcode `\#12\catcode `\^12\catcode `\_12\catcode `\%12\relax}%
\providecommand \@@startlink[1]{}%
\providecommand \@@endlink[0]{}%
\providecommand \url  [0]{\begingroup\@sanitize@url \@url }%
\providecommand \@url [1]{\endgroup\@href {#1}{\urlprefix }}%
\providecommand \urlprefix  [0]{URL }%
\providecommand \Eprint [0]{\href }%
\providecommand \doibase [0]{https://doi.org/}%
\providecommand \selectlanguage [0]{\@gobble}%
\providecommand \bibinfo  [0]{\@secondoftwo}%
\providecommand \bibfield  [0]{\@secondoftwo}%
\providecommand \translation [1]{[#1]}%
\providecommand \BibitemOpen [0]{}%
\providecommand \bibitemStop [0]{}%
\providecommand \bibitemNoStop [0]{.\EOS\space}%
\providecommand \EOS [0]{\spacefactor3000\relax}%
\providecommand \BibitemShut  [1]{\csname bibitem#1\endcsname}%
\let\auto@bib@innerbib\@empty
%</preamble>
\bibitem [{\citenamefont {Hutter}\ \emph {et~al.}(1995)\citenamefont {Hutter},
  \citenamefont {Koch}, \citenamefont {Pluüss},\ and\ \citenamefont
  {Savage}}]{1995The}%
  \BibitemOpen
  \bibfield  {author} {\bibinfo {author} {\bibfnamefont {K.}~\bibnamefont
  {Hutter}}, \bibinfo {author} {\bibfnamefont {T.}~\bibnamefont {Koch}},
  \bibinfo {author} {\bibfnamefont {C.}~\bibnamefont {Pluüss}},\ and\ \bibinfo
  {author} {\bibfnamefont {S.~B.}\ \bibnamefont {Savage}},\ }\bibfield  {title}
  {\bibinfo {title} {The dynamics of avalanches of granular materials from
  initiation to runout. part ii. experiments},\ }\href@noop {} {\bibfield
  {journal} {\bibinfo  {journal} {Acta Mechanica}\ }\textbf {\bibinfo {volume}
  {109}},\ \bibinfo {pages} {127} (\bibinfo {year} {1995})}\BibitemShut
  {NoStop}%
\bibitem [{\citenamefont {Tegzes}\ \emph {et~al.}(2002)\citenamefont {Tegzes},
  \citenamefont {Vicsek},\ and\ \citenamefont {Schiffer}}]{2002Avalanche}%
  \BibitemOpen
  \bibfield  {author} {\bibinfo {author} {\bibfnamefont {P.}~\bibnamefont
  {Tegzes}}, \bibinfo {author} {\bibfnamefont {T.}~\bibnamefont {Vicsek}},\
  and\ \bibinfo {author} {\bibfnamefont {P.}~\bibnamefont {Schiffer}},\
  }\bibfield  {title} {\bibinfo {title} {Avalanche dynamics in wet granular
  materials},\ }\href@noop {} {\bibfield  {journal} {\bibinfo  {journal}
  {Physical Review Letters}\ }\textbf {\bibinfo {volume} {89}},\ \bibinfo
  {pages} {094301} (\bibinfo {year} {2002})}\BibitemShut {NoStop}%
\bibitem [{\citenamefont {Yang}\ \emph {et~al.}(2020)\citenamefont {Yang},
  \citenamefont {Jing}, \citenamefont {Kwok},\ and\ \citenamefont
  {Sobral}}]{2020Pore}%
  \BibitemOpen
  \bibfield  {author} {\bibinfo {author} {\bibfnamefont {G.~C.}\ \bibnamefont
  {Yang}}, \bibinfo {author} {\bibfnamefont {L.}~\bibnamefont {Jing}}, \bibinfo
  {author} {\bibfnamefont {C.~Y.}\ \bibnamefont {Kwok}},\ and\ \bibinfo
  {author} {\bibfnamefont {Y.~D.}\ \bibnamefont {Sobral}},\ }\bibfield  {title}
  {\bibinfo {title} {Pore‐scale simulation of immersed granular collapse:
  Implications to submarine landslides},\ }\href@noop {} {\bibfield  {journal}
  {\bibinfo  {journal} {Journal of Geophysical Research: Earth Surface}\
  }\textbf {\bibinfo {volume} {125}} (\bibinfo {year} {2020})}\BibitemShut
  {NoStop}%
\bibitem [{\citenamefont {Iordanoff}\ and\ \citenamefont
  {Khonsari}(2004)}]{iordanoff2004granular}%
  \BibitemOpen
  \bibfield  {author} {\bibinfo {author} {\bibfnamefont {I.}~\bibnamefont
  {Iordanoff}}\ and\ \bibinfo {author} {\bibfnamefont {M.}~\bibnamefont
  {Khonsari}},\ }\bibfield  {title} {\bibinfo {title} {Granular lubrication:
  toward an understanding of the transition between kinetic and quasi-fluid
  regime},\ }\href@noop {} {\bibfield  {journal} {\bibinfo  {journal} {J.
  Trib.}\ }\textbf {\bibinfo {volume} {126}},\ \bibinfo {pages} {137} (\bibinfo
  {year} {2004})}\BibitemShut {NoStop}%
\bibitem [{\citenamefont {Da~Cruz}\ \emph {et~al.}(2005)\citenamefont
  {Da~Cruz}, \citenamefont {Emam}, \citenamefont {Prochnow}, \citenamefont
  {Roux},\ and\ \citenamefont {Chevoir}}]{da2005rheophysics}%
  \BibitemOpen
  \bibfield  {author} {\bibinfo {author} {\bibfnamefont {F.}~\bibnamefont
  {Da~Cruz}}, \bibinfo {author} {\bibfnamefont {S.}~\bibnamefont {Emam}},
  \bibinfo {author} {\bibfnamefont {M.}~\bibnamefont {Prochnow}}, \bibinfo
  {author} {\bibfnamefont {J.-N.}\ \bibnamefont {Roux}},\ and\ \bibinfo
  {author} {\bibfnamefont {F.}~\bibnamefont {Chevoir}},\ }\bibfield  {title}
  {\bibinfo {title} {Rheophysics of dense granular materials: Discrete
  simulation of plane shear flows},\ }\href@noop {} {\bibfield  {journal}
  {\bibinfo  {journal} {Physical Review E}\ }\textbf {\bibinfo {volume} {72}},\
  \bibinfo {pages} {021309} (\bibinfo {year} {2005})}\BibitemShut {NoStop}%
\bibitem [{\citenamefont {Jop}\ \emph {et~al.}(2006)\citenamefont {Jop},
  \citenamefont {Forterre},\ and\ \citenamefont
  {Pouliquen}}]{jop2006constitutive}%
  \BibitemOpen
  \bibfield  {author} {\bibinfo {author} {\bibfnamefont {P.}~\bibnamefont
  {Jop}}, \bibinfo {author} {\bibfnamefont {Y.}~\bibnamefont {Forterre}},\ and\
  \bibinfo {author} {\bibfnamefont {O.}~\bibnamefont {Pouliquen}},\ }\bibfield
  {title} {\bibinfo {title} {A constitutive law for dense granular flows},\
  }\href@noop {} {\bibfield  {journal} {\bibinfo  {journal} {Nature}\ }\textbf
  {\bibinfo {volume} {441}},\ \bibinfo {pages} {727} (\bibinfo {year}
  {2006})}\BibitemShut {NoStop}%
\bibitem [{\citenamefont {GDR}\ and\ \citenamefont {MiDi}(2004)}]{GDR2004On}%
  \BibitemOpen
  \bibfield  {author} {\bibinfo {author} {\bibnamefont {GDR}}\ and\ \bibinfo
  {author} {\bibnamefont {MiDi}},\ }\bibfield  {title} {\bibinfo {title} {On
  dense granular flows},\ }\href@noop {} {\bibfield  {journal} {\bibinfo
  {journal} {European Physical Journal E}\ } (\bibinfo {year}
  {2004})}\BibitemShut {NoStop}%
\bibitem [{\citenamefont {Lacaze}\ and\ \citenamefont
  {Kerswell}(2009)}]{lacaze2009axisymmetric}%
  \BibitemOpen
  \bibfield  {author} {\bibinfo {author} {\bibfnamefont {L.}~\bibnamefont
  {Lacaze}}\ and\ \bibinfo {author} {\bibfnamefont {R.~R.}\ \bibnamefont
  {Kerswell}},\ }\bibfield  {title} {\bibinfo {title} {Axisymmetric granular
  collapse: a transient 3d flow test of viscoplasticity},\ }\href@noop {}
  {\bibfield  {journal} {\bibinfo  {journal} {Physical Review Letters}\
  }\textbf {\bibinfo {volume} {102}},\ \bibinfo {pages} {108305} (\bibinfo
  {year} {2009})}\BibitemShut {NoStop}%
\bibitem [{\citenamefont {Boyer}\ \emph {et~al.}(2011)\citenamefont {Boyer},
  \citenamefont {Guazzelli},\ and\ \citenamefont
  {Pouliquen}}]{boyer2011unifying}%
  \BibitemOpen
  \bibfield  {author} {\bibinfo {author} {\bibfnamefont {F.}~\bibnamefont
  {Boyer}}, \bibinfo {author} {\bibfnamefont {{\'E}.}~\bibnamefont
  {Guazzelli}},\ and\ \bibinfo {author} {\bibfnamefont {O.}~\bibnamefont
  {Pouliquen}},\ }\bibfield  {title} {\bibinfo {title} {Unifying suspension and
  granular rheology},\ }\href@noop {} {\bibfield  {journal} {\bibinfo
  {journal} {Physical review letters}\ }\textbf {\bibinfo {volume} {107}},\
  \bibinfo {pages} {188301} (\bibinfo {year} {2011})}\BibitemShut {NoStop}%
\bibitem [{\citenamefont {Trulsson}\ \emph {et~al.}(2012)\citenamefont
  {Trulsson}, \citenamefont {Andreotti},\ and\ \citenamefont
  {Claudin}}]{trulsson2012transition}%
  \BibitemOpen
  \bibfield  {author} {\bibinfo {author} {\bibfnamefont {M.}~\bibnamefont
  {Trulsson}}, \bibinfo {author} {\bibfnamefont {B.}~\bibnamefont
  {Andreotti}},\ and\ \bibinfo {author} {\bibfnamefont {P.}~\bibnamefont
  {Claudin}},\ }\bibfield  {title} {\bibinfo {title} {Transition from the
  viscous to inertial regime in dense suspensions},\ }\href@noop {} {\bibfield
  {journal} {\bibinfo  {journal} {Physical review letters}\ }\textbf {\bibinfo
  {volume} {109}},\ \bibinfo {pages} {118305} (\bibinfo {year}
  {2012})}\BibitemShut {NoStop}%
\bibitem [{\citenamefont {Lacaze}\ \emph {et~al.}(2021)\citenamefont {Lacaze},
  \citenamefont {Bouteloup}, \citenamefont {Fry},\ and\ \citenamefont
  {Izard}}]{lacaze2021immersed}%
  \BibitemOpen
  \bibfield  {author} {\bibinfo {author} {\bibfnamefont {L.}~\bibnamefont
  {Lacaze}}, \bibinfo {author} {\bibfnamefont {J.}~\bibnamefont {Bouteloup}},
  \bibinfo {author} {\bibfnamefont {B.}~\bibnamefont {Fry}},\ and\ \bibinfo
  {author} {\bibfnamefont {E.}~\bibnamefont {Izard}},\ }\bibfield  {title}
  {\bibinfo {title} {Immersed granular collapse: from viscous to free-fall
  unsteady granular flows},\ }\href@noop {} {\bibfield  {journal} {\bibinfo
  {journal} {Journal of Fluid Mechanics}\ }\textbf {\bibinfo {volume} {912}}
  (\bibinfo {year} {2021})}\BibitemShut {NoStop}%
\bibitem [{\citenamefont {Du~Pont}\ \emph {et~al.}(2003)\citenamefont
  {Du~Pont}, \citenamefont {Gondret}, \citenamefont {Perrin},\ and\
  \citenamefont {Rabaud}}]{du2003granular}%
  \BibitemOpen
  \bibfield  {author} {\bibinfo {author} {\bibfnamefont {S.~C.}\ \bibnamefont
  {Du~Pont}}, \bibinfo {author} {\bibfnamefont {P.}~\bibnamefont {Gondret}},
  \bibinfo {author} {\bibfnamefont {B.}~\bibnamefont {Perrin}},\ and\ \bibinfo
  {author} {\bibfnamefont {M.}~\bibnamefont {Rabaud}},\ }\bibfield  {title}
  {\bibinfo {title} {Granular avalanches in fluids},\ }\href@noop {} {\bibfield
   {journal} {\bibinfo  {journal} {Physical review letters}\ }\textbf {\bibinfo
  {volume} {90}},\ \bibinfo {pages} {044301} (\bibinfo {year}
  {2003})}\BibitemShut {NoStop}%
\bibitem [{\citenamefont {Forterre}\ and\ \citenamefont
  {Pouliquen}(2008)}]{forterre2008flows}%
  \BibitemOpen
  \bibfield  {author} {\bibinfo {author} {\bibfnamefont {Y.}~\bibnamefont
  {Forterre}}\ and\ \bibinfo {author} {\bibfnamefont {O.}~\bibnamefont
  {Pouliquen}},\ }\bibfield  {title} {\bibinfo {title} {Flows of dense granular
  media},\ }\href@noop {} {\bibfield  {journal} {\bibinfo  {journal} {Annu.
  Rev. Fluid Mech.}\ }\textbf {\bibinfo {volume} {40}},\ \bibinfo {pages} {1}
  (\bibinfo {year} {2008})}\BibitemShut {NoStop}%
\bibitem [{\citenamefont {Cundall}\ and\ \citenamefont
  {Strack}(1979)}]{cundall1979discrete}%
  \BibitemOpen
  \bibfield  {author} {\bibinfo {author} {\bibfnamefont {P.~A.}\ \bibnamefont
  {Cundall}}\ and\ \bibinfo {author} {\bibfnamefont {O.~D.}\ \bibnamefont
  {Strack}},\ }\bibfield  {title} {\bibinfo {title} {A discrete numerical model
  for granular assemblies},\ }\href@noop {} {\bibfield  {journal} {\bibinfo
  {journal} {geotechnique}\ }\textbf {\bibinfo {volume} {29}},\ \bibinfo
  {pages} {47} (\bibinfo {year} {1979})}\BibitemShut {NoStop}%
\bibitem [{\citenamefont {Galindo-Torres}(2013)}]{galindo2013coupled}%
  \BibitemOpen
  \bibfield  {author} {\bibinfo {author} {\bibfnamefont {S.}~\bibnamefont
  {Galindo-Torres}},\ }\bibfield  {title} {\bibinfo {title} {A coupled discrete
  element lattice boltzmann method for the simulation of fluid--solid
  interaction with particles of general shapes},\ }\href@noop {} {\bibfield
  {journal} {\bibinfo  {journal} {Computer Methods in Applied Mechanics and
  Engineering}\ }\textbf {\bibinfo {volume} {265}},\ \bibinfo {pages} {107}
  (\bibinfo {year} {2013})}\BibitemShut {NoStop}%
\bibitem [{\citenamefont {Bougouin}\ and\ \citenamefont
  {Lacaze}(2018)}]{bougouin2018granular}%
  \BibitemOpen
  \bibfield  {author} {\bibinfo {author} {\bibfnamefont {A.}~\bibnamefont
  {Bougouin}}\ and\ \bibinfo {author} {\bibfnamefont {L.}~\bibnamefont
  {Lacaze}},\ }\bibfield  {title} {\bibinfo {title} {Granular collapse in a
  fluid: Different flow regimes for an initially dense-packing},\ }\href@noop
  {} {\bibfield  {journal} {\bibinfo  {journal} {Physical Review Fluids}\
  }\textbf {\bibinfo {volume} {3}},\ \bibinfo {pages} {064305} (\bibinfo {year}
  {2018})}\BibitemShut {NoStop}%
\bibitem [{\citenamefont {Yan}\ and\ \citenamefont
  {Regueiro}(2019)}]{yan2019definition}%
  \BibitemOpen
  \bibfield  {author} {\bibinfo {author} {\bibfnamefont {B.}~\bibnamefont
  {Yan}}\ and\ \bibinfo {author} {\bibfnamefont {R.~A.}\ \bibnamefont
  {Regueiro}},\ }\bibfield  {title} {\bibinfo {title} {Definition and symmetry
  of averaged stress tensor in granular media and its 3d dem inspection under
  static and dynamic conditions},\ }\href@noop {} {\bibfield  {journal}
  {\bibinfo  {journal} {International Journal of Solids and Structures}\
  }\textbf {\bibinfo {volume} {161}},\ \bibinfo {pages} {243} (\bibinfo {year}
  {2019})}\BibitemShut {NoStop}%
\bibitem [{\citenamefont {Goldhirsch}\ and\ \citenamefont
  {Goldenberg}(2002)}]{goldhirsch2002microscopic}%
  \BibitemOpen
  \bibfield  {author} {\bibinfo {author} {\bibfnamefont {I.}~\bibnamefont
  {Goldhirsch}}\ and\ \bibinfo {author} {\bibfnamefont {C.}~\bibnamefont
  {Goldenberg}},\ }\bibfield  {title} {\bibinfo {title} {On the microscopic
  foundations of elasticity},\ }\href@noop {} {\bibfield  {journal} {\bibinfo
  {journal} {The European Physical Journal E}\ }\textbf {\bibinfo {volume}
  {9}},\ \bibinfo {pages} {245} (\bibinfo {year} {2002})}\BibitemShut {NoStop}%
\bibitem [{\citenamefont {Quartier}\ \emph {et~al.}(2000)\citenamefont
  {Quartier}, \citenamefont {Andreotti}, \citenamefont {Douady},\ and\
  \citenamefont {Daerr}}]{quartier2000dynamics}%
  \BibitemOpen
  \bibfield  {author} {\bibinfo {author} {\bibfnamefont {L.}~\bibnamefont
  {Quartier}}, \bibinfo {author} {\bibfnamefont {B.}~\bibnamefont {Andreotti}},
  \bibinfo {author} {\bibfnamefont {S.}~\bibnamefont {Douady}},\ and\ \bibinfo
  {author} {\bibfnamefont {A.}~\bibnamefont {Daerr}},\ }\bibfield  {title}
  {\bibinfo {title} {Dynamics of a grain on a sandpile model},\ }\href@noop {}
  {\bibfield  {journal} {\bibinfo  {journal} {Physical Review E}\ }\textbf
  {\bibinfo {volume} {62}},\ \bibinfo {pages} {8299} (\bibinfo {year}
  {2000})}\BibitemShut {NoStop}%
\bibitem [{\citenamefont {Kim}\ and\ \citenamefont
  {Kamrin}(2020)}]{kim2020power}%
  \BibitemOpen
  \bibfield  {author} {\bibinfo {author} {\bibfnamefont {S.}~\bibnamefont
  {Kim}}\ and\ \bibinfo {author} {\bibfnamefont {K.}~\bibnamefont {Kamrin}},\
  }\bibfield  {title} {\bibinfo {title} {Power-law scaling in granular rheology
  across flow geometries},\ }\href@noop {} {\bibfield  {journal} {\bibinfo
  {journal} {Physical Review Letters}\ }\textbf {\bibinfo {volume} {125}},\
  \bibinfo {pages} {088002} (\bibinfo {year} {2020})}\BibitemShut {NoStop}%
\bibitem [{\citenamefont {Roux}\ and\ \citenamefont
  {Combe}(2002)}]{roux2002quasistatic}%
  \BibitemOpen
  \bibfield  {author} {\bibinfo {author} {\bibfnamefont {J.-N.}\ \bibnamefont
  {Roux}}\ and\ \bibinfo {author} {\bibfnamefont {G.}~\bibnamefont {Combe}},\
  }\bibfield  {title} {\bibinfo {title} {Quasistatic rheology and the origins
  of strain},\ }\href@noop {} {\bibfield  {journal} {\bibinfo  {journal}
  {Comptes Rendus Physique}\ }\textbf {\bibinfo {volume} {3}},\ \bibinfo
  {pages} {131} (\bibinfo {year} {2002})}\BibitemShut {NoStop}%
\bibitem [{\citenamefont {Goldhirsch}(2003)}]{goldhirsch2003rapid}%
  \BibitemOpen
  \bibfield  {author} {\bibinfo {author} {\bibfnamefont {I.}~\bibnamefont
  {Goldhirsch}},\ }\bibfield  {title} {\bibinfo {title} {Rapid granular
  flows},\ }\href@noop {} {\bibfield  {journal} {\bibinfo  {journal} {Annual
  review of fluid mechanics}\ }\textbf {\bibinfo {volume} {35}},\ \bibinfo
  {pages} {267} (\bibinfo {year} {2003})}\BibitemShut {NoStop}%
\bibitem [{\citenamefont {Cassar}\ \emph {et~al.}(2005)\citenamefont {Cassar},
  \citenamefont {Nicolas},\ and\ \citenamefont
  {Pouliquen}}]{cassar2005submarine}%
  \BibitemOpen
  \bibfield  {author} {\bibinfo {author} {\bibfnamefont {C.}~\bibnamefont
  {Cassar}}, \bibinfo {author} {\bibfnamefont {M.}~\bibnamefont {Nicolas}},\
  and\ \bibinfo {author} {\bibfnamefont {O.}~\bibnamefont {Pouliquen}},\
  }\bibfield  {title} {\bibinfo {title} {Submarine granular flows down inclined
  planes},\ }\href@noop {} {\bibfield  {journal} {\bibinfo  {journal} {Physics
  of fluids}\ }\textbf {\bibinfo {volume} {17}},\ \bibinfo {pages} {103301}
  (\bibinfo {year} {2005})}\BibitemShut {NoStop}%
\bibitem [{\citenamefont {Todd}\ \emph {et~al.}(1995)\citenamefont {Todd},
  \citenamefont {Evans},\ and\ \citenamefont {Daivis}}]{todd1995pressure}%
  \BibitemOpen
  \bibfield  {author} {\bibinfo {author} {\bibfnamefont {B.}~\bibnamefont
  {Todd}}, \bibinfo {author} {\bibfnamefont {D.~J.}\ \bibnamefont {Evans}},\
  and\ \bibinfo {author} {\bibfnamefont {P.~J.}\ \bibnamefont {Daivis}},\
  }\bibfield  {title} {\bibinfo {title} {Pressure tensor for inhomogeneous
  fluids},\ }\href@noop {} {\bibfield  {journal} {\bibinfo  {journal} {Physical
  Review E}\ }\textbf {\bibinfo {volume} {52}},\ \bibinfo {pages} {1627}
  (\bibinfo {year} {1995})}\BibitemShut {NoStop}%
\bibitem [{\citenamefont {Subramaniyan}\ and\ \citenamefont
  {Sun}(2008)}]{subramaniyan2008continuum}%
  \BibitemOpen
  \bibfield  {author} {\bibinfo {author} {\bibfnamefont {A.~K.}\ \bibnamefont
  {Subramaniyan}}\ and\ \bibinfo {author} {\bibfnamefont {C.}~\bibnamefont
  {Sun}},\ }\bibfield  {title} {\bibinfo {title} {Continuum interpretation of
  virial stress in molecular simulations},\ }\href@noop {} {\bibfield
  {journal} {\bibinfo  {journal} {International Journal of Solids and
  Structures}\ }\textbf {\bibinfo {volume} {45}},\ \bibinfo {pages} {4340}
  (\bibinfo {year} {2008})}\BibitemShut {NoStop}%
\end{thebibliography}%
    \section{Appendix}
    
    The settling velocity of a particle in fluid is expressed as
    \begin{equation}
    u(t)=u_{f}(1-e^{-at}),
    \end{equation}
    with $a=18\eta_{f}/(\rho_{s}d^2)$.
    The traveling distance of the particle is
    \begin{equation}
    s_{1}(t)=\int_{0}^{t}u(t)dt=u_{f}t-\frac{u_{f}}{a}(1-e^{-at})
    \end{equation}
    
\begin{figure}
    	\centering
    	\includegraphics[scale=0.22]{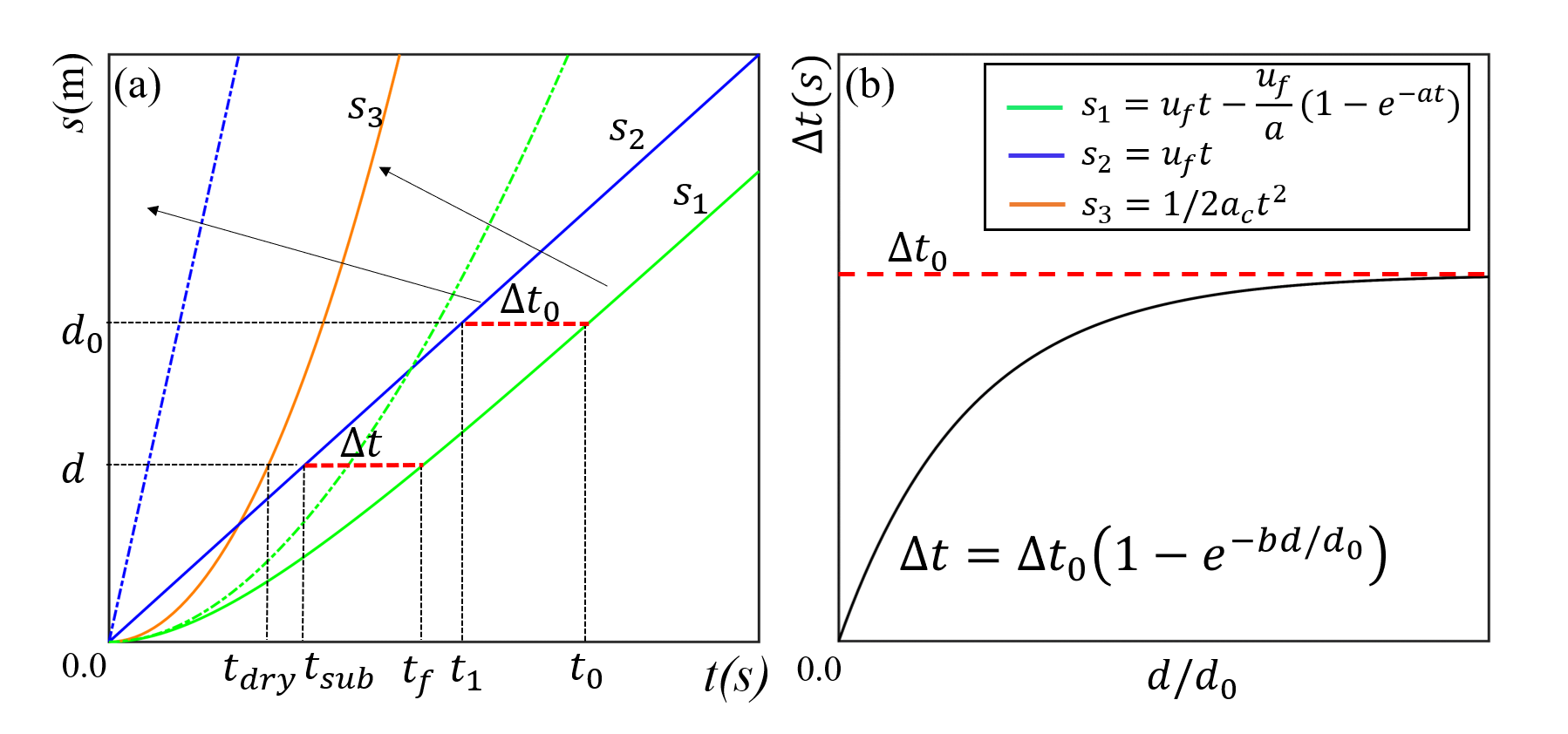}
    	\caption{Single particle settling in the fluid: (a) Evolution of the traveling distance in terms of time; $s_{1}$, particle travels in fluid under the hydrodynamic force; $s_{2}$ , particle traveling with the final velocity $u_{f}=Pd/(12\eta_{f})$, where the inertial force equal to the hydrodynamic force; $s_{3}$ particle traveling with a constant acceleration due to the inertial force. The green and blue dot lines represent that with decreasing of $\eta_{f}$, $s_{1}$ is close to $s_{3}$, $s_{2}$ is close to $t=0$. (b) Variation of time difference between $s_{1}$ and $s_{2}$ in terms of traveling distance $d/d_{0}$. }
    	\label{settle}
    \end{figure}

    The relation between different traveling distances and the time is shown in Fig. \ref{settle}(a), $s_{3}=a_{c}t^{2}/2$ is for the particle settling in dry conditions with the constant acceleration $a_{c}=3P/(2\rho_{s}d)$, $s_{2}=u_{f}t$ is for the particle traveling with the final velocity $u_{f}=Pd/(12\eta_{f})$, which is obtained in the condition of inertial force ($(\pi/4)Pd^{2}$) equaling to the Stokes force ($F_{d}=3\pi\eta_{f}du_{f}$), and $s_{1}$ is for the particle settling in fluid with the Stokes force $F_{d}=3\pi\eta_{f}d u_{p}$. The settling time calculated from different conditions shows that $t_{sub}$ and $t_{dry}$ are always smaller than $t_{f}$. The settling time can also be expressed as $t_{f}=t_{sub}+\Delta t$, where $\Delta t$ is a time gap between the particle settling at terminal velocity and the one starting from rest. 
    As shown in Fig. \ref{settle}(a), increasing the traveling distance, $\Delta t$ increases from 0 to a final time difference $\Delta t_{0}$. We assume that, when the settling velocity $u=\alpha u_{f}$, where $\alpha \approx 1$, the time difference is not changing anymore. In this case, the settling time $t_{0}=-\ln(1-\alpha)/a$, and the traveling distance is $d_{0}=-u_{f}/a[\ln(1-\alpha)+\alpha]$. Due to $t_{1}=d_{0}/u_{f}$, the final time difference $\Delta t_{0}=t_{0}-t_{1}=\alpha/a$. Hence, the time difference could be given by $\Delta t=\Delta t_{0}(1-e^{-bd/d_{0}})$. We can obtain
    \begin{equation}
    \begin{split}
        \frac{d}{d_{0}}&=-\frac{ad}{u_{f}[\ln(1-\alpha)+\alpha]}\\
        &=-\frac{18\eta_{f} d}{\rho_{s}d^{2}}\frac{12\eta_{f}}{Pd[\ln{(1-\alpha)}+\alpha]}\\
    \end{split}
    \end{equation} 
    and the settling time of the particle in the fluid is
    \begin{equation}
        t_{f}=\frac{12\eta_{f}}{P}+\frac{\rho_{s}d^2\alpha}{18\eta_{f}}\left[1-e^{\frac{216\eta_{f}^2}{P \rho_{s} d^2}\frac{b}{\ln(1-\alpha)+\alpha}} \right].\label{tf}
    \end{equation}
    Due to $\lim_{\eta_{f} \to 0} t_{f}=t_{dry}=\sqrt{4\rho_{s}d^2/(3P)}$, one can obtain $\frac{b}{\ln(1-\alpha)+\alpha}=-\frac{\sqrt{\rho_{s}P}d}{6\sqrt{3}\eta_{f}\alpha}$, and Eq.\ref{tf} is 
    \begin{equation}
        t_{f}=\frac{12\eta_{f}}{P}+\frac{\rho_{s}d^2\alpha}{18\eta_{f}}\left [1-e^{-\frac{36\eta_{f}}{\sqrt{3P\rho_{s}}d\alpha}}\right]
    \end{equation}
    
    Hence, the ratio of the microscopic particle time scale to the macroscopic rearrangement time scale is given by
    \begin{equation}
    \begin{split}
            \mathcal{G}&=\frac{t_{f}}{\mathcal{T}}=\frac{12\eta_{f}\dot{\gamma}}{P}+\frac{\alpha\rho_{s}d^2\dot{\gamma}}{18\eta_{f}}\left [1-e^{-\frac{36\eta_{f}}{\sqrt{3P\rho_{s}}d\alpha}}\right]\\
        &=\frac{12\dot{\gamma}\eta_{f}}{P}+\frac{\alpha P}{18\eta_{f}\dot{\gamma}}\frac{\rho_{s}d^2\dot{\gamma}^2}{P}\left[1-e^{-\frac{36}{\sqrt{3}}\frac{\eta_{f}}{P}\frac{\sqrt{P}}{\sqrt{\rho_{s}d\alpha}}}\right]\\
        &=12I_{v}+\frac{\alpha I^2}{18 I_{v}}\left [1-e^{-\frac{36}{\alpha\sqrt{3}}\frac{I_{v}}{I}}\right ],
    \end{split}
    \label{eq:G}
    \end{equation}
    
    since $\alpha$ is close to 1, and the dimensionless number $\mathcal{G}$ could be expressed as
    \begin{equation}
    \begin{split}
        \mathcal{G}&=12I_{v}+\frac{\alpha I^2}{18 I_{v}}\left [1-e^{-\frac{36}{\alpha\sqrt{3}}\frac{I_{v}}{I}}\right ]\\
        &=I_{v}\left [12+\frac{\mathcal{ST}^2}{18}-\frac{\mathcal{ST}^2}{18}e^{-\frac{36}{\sqrt{3}\mathcal{ST}}}\right ]\\
        &=\frac{I}{\mathcal{ST}}\left [12+\frac{\mathcal{ST}^2}{18}-\frac{\mathcal{ST}^2}{18}e^{-\frac{36}{\sqrt{3}\mathcal{ST}}}\right ].
    \end{split}
    \end{equation}
    When the inertial force is dominant, the hydrodynamic force is nearly nil, one can obtain
    \begin{equation}
    \begin{split}
            \lim_{\mathcal{ST}\to +\infty}\mathcal{G}&=\frac{I}{\mathcal{ST}}\left [12+\frac{\mathcal{ST}^2}{18}-\frac{\mathcal{ST}^2}{18}e^{-\frac{36}{\sqrt{3}\mathcal{ST}}}\right ]\\
            &=\frac{I}{18}\left [1-e^{-\frac{36}{\sqrt{3}\mathcal{ST}}}\right]\mathcal{ST}\\
            &\approx \frac{I}{18}\left [ \frac{36}{\sqrt{3}\mathcal{ST}}\right]\mathcal{ST}\\
            &=\frac{2}{\sqrt{3}}I=\frac{t_{dry}}{\mathcal{T}},    
    \end{split}
    \end{equation}
    when the hydrodynamic force is dominant, one can obtain
    \begin{equation}
    \begin{split}
    \lim_{\mathcal{ST}\to +0}\mathcal{G}&=I_{v}\left [12+\frac{\mathcal{ST}^2}{18}-\frac{\mathcal{ST}^2}{18}e^{-\frac{36}{\sqrt{3}\mathcal{ST}}}\right]\\
    &=12I_{v}=\frac{t_{sub}}{\mathcal{T}}
    \end{split}
    \end{equation}
\end{document}